\documentclass[11pt,a4paper,pdftex]{article}
\pdfoutput=1
\usepackage{jcappub}

\newcommand{\be}{\begin{eqnarray}}
\newcommand{\ee}{\end{eqnarray}}

\begin{document}

\title{Measuring the Kerr spin parameter of a non-Kerr compact object with the continuum-fitting and the iron line methods}

\author{Cosimo Bambi}
\emailAdd{bambi@fudan.edu.cn}

\affiliation{Center for Field Theory and Particle Physics \& Department of Physics,\\
Fudan University, 200433 Shanghai, China}

\abstract{Under the assumption that astrophysical black hole candidates are the 
Kerr black holes of general relativity, the continuum-fitting method and the analysis 
of the K$\alpha$ iron line are today the only available techniques capable of 
providing a relatively reliable estimate of the spin parameter of these objects. 
If we relax the Kerr black hole hypothesis and we try to test the nature of black 
hole candidates, we find that there is a strong correlation between the measurement
of the spin and possible deviations from the Kerr solution. The properties of the
radiation emitted in a Kerr spacetime with spin parameter $a_*$ are indeed
very similar, and practically indistinguishable, from the ones of the radiation
emitted around a non-Kerr object with different spin. In this paper, I address the 
question whether measuring the Kerr spin with both the continuum-fitting method 
and the K$\alpha$ iron line analysis of the same object can be used to claim the 
Kerr nature of the black hole candidate in the case of consistent results. In this
work, I consider two non-Kerr metrics and it seems that the answer does depend
on the specific background. The two techniques may either provide a very similar 
result (the case of the Bardeen metric) or show a discrepancy (Johannsen-Psaltis
background).}


\maketitle


\section{Introduction}

Today we have robust observational evidence for the existence of compact 
objects in X-ray binary systems with a mass $M \approx 5 - 20$~$M_\odot$,
and of dark bodies at the center of every normal galaxy with
$M \sim 10^5 - 10^9$~$M_\odot$ (for a review, see e.g. Ref.~\cite{bh1}). 
The masses of these objects are inferred by dynamical measurements, by
studying the Newtonian orbital motion of gas or individual stars, and their 
estimate is thus reliable. It is thought that all these objects are the Kerr black 
holes (BHs) of general relativity. Stellar-mass BH candidates are indeed 
too heavy to be neutron or quark stars for any reasonable matter equation 
of state~\cite{bh2-a,bh2-b}. The super-massive BH candidates at the centers of 
galaxies turn out to be too massive, compact, and old to be clusters of
non-luminous objects, as the cluster lifetime due to evaporation and physical 
collisions would be shorter than the age of these systems~\cite{bh3}. The 
non-observation of thermal radiation emitted by the possible surface of 
BH candidates may also be interpreted as an indication for the existence of 
an event horizon~\cite{bh4-a,bh4-b} (see however~\cite{horizon-a,horizon-b}). 
But that is all we know. There is no direct evidence that BH candidates are 
Kerr BHs and current observations cannot rule out the possibility that the 
spacetime geometry around them significantly deviates from the Kerr 
solution~\cite{review-a,review-b}.

In 4-dimensional general relativity, uncharged BHs are described by the
Kerr metric and they are completely specified by only two parameters: the
mass $M$ and the spin parameter $a_* = J/M^2$, where $J$ is the BH spin
angular momentum~\cite{kerr-a,kerr-b,kerr-c}. The condition for the existence of the event 
horizon is $|a_*| \le 1$, while for $|a_*| > 1$ there is no BH and the central 
singularity is naked. All the properties of the spacetime can be immediately 
deduced from $M$ and $a_*$. If astrophysical BH candidates are the Kerr BHs 
of general relativity, we just need to measure their mass and spin parameter 
to know everything about them. The mass $M$ is relatively easy to obtain,
as it can be inferred by studying the orbital motion of individual stars around 
the BH candidate in the framework of Newtonian mechanics; that is, without
any assumption about the nature of the compact object. The measurement 
of the spin is much more challenging. The spin has indeed no gravitational
effects in Newtonian mechanics and thus its measurement requires to
probe the spacetime close to the compact object. That is possible, at least
in principle, by studying the properties of the electromagnetic radiation 
emitted by the gas in the accretion disk.

At present, there are only two techniques capable of providing a relatively
reliable measurement of the spin parameter of BH candidates under the
assumption that the spacetime around them is described by the Kerr
solution: the continuum-fitting method~\cite{cfm-a,cfm-b,cfm-c} and the analysis of the
K$\alpha$ iron line~\cite{ka-a,ka-b}. With the continuum-fitting method, one studies 
the thermal spectrum of geometrically thin and optically thick accretion disks
and the spin parameter can be inferred from the position of the high energy 
cut-off. In the case of the iron line, one considers a line that is intrinsically
narrow in frequency, while the one observed appears broadened and 
skewed due to special and general relativistic effects. Here the spin 
parameter is measured from the low energy tail of the line. Let us note that
the continuum-fitting method can be applied only to stellar-mass BH
candidates: the disk's temperature scales as $M^{-0.25}$, so the high 
energy part of the spectrum is in the keV band for stellar-mass BH candidates 
and in the UV range for super-massive ones. In the latter case, dust 
absorption makes a measurement impossible. The iron line approach has not
this problem, and it can be used to measure the spin of both stellar-mass 
and super-massive BH candidates.

Both the continuum-fitting method and the analysis of the K$\alpha$ iron line can 
be extended to non-Kerr backgrounds and be used to test the nature of BH 
candidates~\cite{harko-a,harko-b,harko-c,cfm-cb-a,cfm-cb-b,ka-cb-a,ka-cb-b,ka-jp,ka-dan}. 
In this case, we cannot measure the spin parameter of the compact object, but 
only constrain a certain combination of the spin parameter and of possible
deviations from the Kerr background. In other words, the thermal
spectrum of thin disks and the shape of the K$\alpha$ iron line of 
Kerr BHs can be very similar, and practically indistinguishable, from
the ones of a non-Kerr compact object with very different spin parameter.

How about we combine the two measurements? Can we break the degeneracy
in the spin-deformation plane and get a measurement for both the spin 
and the deformations? In general, it is not easy to measure the
spin of a stellar-mass BH candidate with both the continuum-fitting method 
and the iron line approach, as the former requires good observational data 
of the soft X-ray component, when the BH candidate is in the high-soft state,
while the latter typically requires that the object is in the low-hard state.
However, it is definitively possible in principle. The measurement of the spin
parameter with both the techniques has been already done in
Ref.~\cite{xte} for the BH candidate XTE~J1550-564, and it may become
a routine analysis in the future, when more data will be available.
To address this question, in this paper I study two different non-Kerr
BH metrics: the Bardeen metric~\cite{regular} and the Johannsen-Psaltis 
one~\cite{jp-m}. While deformations of the Bardeen type seem to be
more theoretically motivated, there may be theories with BHs similar
to the Johannsen-Psaltis ones. The results of the present work show that 
the continuum-fitting and the iron line measurements, when combined 
and if consistent, cannot break this degeneracy in the Bardeen case,
but they can potentially do it for Johannsen-Psaltis BHs. So, some caution
should be in order before claiming that astrophysical BH candidates are 
the Kerr BHs of general relativity in the case in which the two approaches
give very similar values of the Kerr spin, even for very accurate measurements.

The paper is organized as follows. In Sections~\ref{s-cfm} and \ref{s-ka}, I 
briefly review, respectively, the continuum-fitting method and the K$\alpha$
iron line analysis. Section~\ref{s-test} is devoted to the use of these two 
techniques to test the nature of BH candidates and I compute their Kerr spin 
measurements in the case BH candidates are not Kerr BHs. In Section~\ref{s-d}, 
I discuss these results. The conclusion is that the two approaches may
be consistent even in the case of non-Kerr BHs, so some caution is definitively
necessary. Summary and conclusions are reported in Section~\ref{s-c}. 
Throughout the manuscript, I use units in which $G_{\rm N} = c = 1$, unless 
stated otherwise.

\section{Continuum-fitting method \label{s-cfm}}

Geometrically thin and optically thick accretion disks around BHs can
be described by the Novikov-Thorne model~\cite{nt}. In a generic
stationary, axisymmetric, and asymptotically flat spacetime, one assumes
that the disk is on the equatorial plane, that the disk's gas moves on 
nearly geodesic circular orbits, and that the energy is radiated from the 
disk's surface. From the conservation laws for the rest-mass, the angular
momentum, and the energy, one obtains three basic equations for the 
time-averaged radial structure of the disk~\cite{pt}. The time-averaged
energy flux is given by
\be
\mathcal{F}(r) = \frac{\dot{M}}{4 \pi \sqrt{-G}} F(r) \, ,
\ee
where $\dot{M}$ is the mass accretion rate, $G$ is the determinant of
the near equatorial plane metric in cylindrical coordinates, so 
$\sqrt{-G} = \sqrt{\alpha^2 g_{rr} g_{\phi\phi}}$ and $\alpha$ is 
the lapse function, and $F(r)$ is
\be\label{eq-f}
F(r) = \frac{\partial_r \Omega}{(E - \Omega L_z)^2} \int_{r_{\rm in}}^r
(E - \Omega L_z)(\partial_\rho L_z) d\rho \, .
\ee
In Eq.~\eqref{eq-f}, $E$, $L_z$, and $\Omega$ are, respectively, the
conserved specific energy, the conserved $z$-component of the specific
angular momentum, and the angular velocity for equatorial circular 
geodesics. $r_{\rm in}$ is the inner radius of the accretion disk and in the
Novikov-Thorne model is assumed to be at the innermost stable circular 
orbit (ISCO).

Since the disk is in thermal equilibrium, the emission is blackbody-like 
and we can define an effective temperature $T_{\rm eff} (r)$ from the 
relation $\mathcal{F}(r) = \sigma T^4_{\rm eff}$, where $\sigma$ is the 
Stefan-Boltzmann constant. Actually, the disk's temperature near the inner 
edge of the disk can be high, up to $\sim 10^7$~K for stellar-mass BH 
candidates, and non-thermal effects are non-negligible. That is usually 
taken into account by introducing the color factor (or hardening factor) 
$f_{\rm col}$. The color temperature is $T_{\rm col} (r) = f_{\rm col} 
T_{\rm eff}$ and the local specific intensity of the radiation emitted by 
the disk is
\be\label{eq-i-bb}
I_{\rm e}(\nu_{\rm e}) = \frac{2 h \nu^3_{\rm e}}{f_{\rm col}^4} 
\frac{\Upsilon}{\exp\left(\frac{h \nu_{\rm e}}{k_{\rm B} T_{\rm col}}\right) - 1} \, ,
\ee
where $\nu_{\rm e}$ is the photon frequency, $h$ is the Planck's constant, 
$k_{\rm B}$ is the Boltzmann constant, and $\Upsilon$ is a function of the 
angle between the wavevector of the photon emitted by the disk and the 
normal of the disk surface, say $\xi$. The two most common options are 
$\Upsilon = 1$ (isotropic emission) and $\Upsilon = \frac{1}{2} + \frac{3}{4} \cos\xi$ 
(limb-darkened emission).

The calculation of the thermal spectrum of a thin accretion disk has been 
extensively discussed in the literature; see e.g.~\cite{cfm-b,cfm-cb-a} and references
therein. The spectrum can be conveniently written in terms of the photon flux 
number density as measured by a distant observer, $N_{E_{\rm obs}}$. 
In order to include all the relativistic effects, it is necessary to compute the
photon trajectories from the disk, where the photon is emitted, to the image plane
of the distant observer, where it is detected. The photon flux number density 
is eventually given by 
\be\label{eq-n2}
N_{E_{\rm obs}} &=&
\frac{1}{E_{\rm obs}} \int I_{\rm obs}(\nu) d \Omega_{\rm obs} = 
\frac{1}{E_{\rm obs}} \int w^3 I_{\rm e}(\nu_{\rm e}) d \Omega_{\rm obs} = 
\nonumber\\ &=& 
A_1 \left(\frac{E_{\rm obs}}{\rm keV}\right)^2
\int \frac{1}{M^2} \frac{\Upsilon dXdY}{\exp\left[\frac{A_2}{w F^{1/4}} 
\left(\frac{E_{\rm obs}}{\rm keV}\right)\right] - 1} \, ,
\ee
where $I_{\rm obs}$, $E_{\rm obs}$, and $\nu$ are, respectively, the specific
intensity of the radiation, the photon energy, and the photon frequency measured
by the distant observer, $X$ and $Y$ are the coordinates of the position of 
the photon on the sky, as seen by the distant observer, and $d\Omega_{\rm obs} 
= dX dY / D^2$, with $D$ the distance of the source. $w$ is the redshift factor
\be\label{eq-red}
w = \frac{E_{\rm obs}}{E_{\rm e}} = \frac{\nu}{\nu_{\rm e}} = 
\frac{k_\alpha u^{\alpha}_{\rm obs}}{k_\beta u^{\beta}_{\rm e}}\, ,
\ee
$E_{\rm e} = h \nu_{\rm e}$, $k^\alpha$ is the 4-momentum of the photon, 
$u^{\alpha}_{\rm obs} = (-1,0,0,0)$ is the 4-velocity of the distant observer, and 
$u^{\alpha}_{\rm e} = (u^t_{\rm e},0,0, \Omega u^t_{\rm e})$ is the 4-velocity of 
the emitter. $I_{\rm e}(\nu_{\rm e})/\nu_{\rm e}^3 = I_{\rm obs} (\nu_{\rm obs})/\nu^3$ 
follows from the Liouville's theorem. $A_1$ and $A_2$ are given by (here I show 
explicitly $G_{\rm N}$ and $c$)
\be
A_1 &=&  
\frac{2 \left({\rm keV}\right)^2}{f_{\rm col}^4} 
\left(\frac{G_{\rm N} M}{c^3 h D}\right)^2 = 
\frac{0.07205}{f_{\rm col}^4} 
\left(\frac{M}{M_\odot}\right)^2 
\left(\frac{\rm kpc}{D}\right)^2 \, 
{\rm \gamma \, keV^{-1} \, cm^{-2} \, s^{-1}} \, , \nonumber\\
A_2 &=&  
\left(\frac{\rm keV}{k_{\rm B} f_{\rm col}}\right) 
\left(\frac{G_{\rm N} M}{c^3}\right)^{1/2}
\left(\frac{4 \pi \sigma}{\dot{M}}\right)^{1/4} = 
\frac{0.1331}{f_{\rm col}} 
\left(\frac{\rm 10^{18} \, g \, s^{-1}}{\dot{M}}\right)^{1/4}
\left(\frac{M}{M_\odot}\right)^{1/2} \, .
\ee
Using the normalization condition
$g_{\mu\nu}u^{\mu}_{\rm e}u^{\nu}_{\rm e} = -1$, one finds
\be
u^t_{\rm e} = - \frac{1}{\sqrt{-g_{tt} - 2 g_{t\phi} \Omega - g_{\phi\phi} \Omega^2}} \, ,
\ee
and therefore
\be\label{eq-red-g}
w = \frac{\sqrt{-g_{tt} - 2 g_{t\phi} \Omega - g_{\phi\phi} \Omega^2}}{1 + 
\lambda \Omega} \, ,
\ee
where $\lambda = k_\phi/k_t$ is a constant of the motion along the photon path.
Doppler boosting, gravitational redshift, and frame dragging are entirely encoded 
in the redshift factor $w$.

\section{K$\alpha$ iron line \label{s-ka}}

The X-ray spectrum of both stellar-mass and supermassive BH candidates 
is usually characterized by the presence of a power-law component. This 
feature is commonly interpreted as the inverse Compton scattering of thermal 
photons by electrons in a hot corona above the accretion disk. The geometry 
of the corona is not known and several models have been proposed. Such 
a ``primary component'' irradiates also the accretion disk, producing a 
``reflection component'' in the X-ray spectrum. The illumination of the cold 
disk by the primary component also produces spectral lines by fluorescence. 
The strongest line is the K$\alpha$ iron line at 6.4 keV. This line is intrinsically 
narrow in frequency, while the one observed appears broadened and skewed. 
Especially for some sources, this line is extraordinarily stable, in spite of a 
substantial variability of the continuum. This fact suggests that its shape is 
determined by the geometry of the spacetime around the compact object.

The profile of the K$\alpha$ iron line depends on the background metric, 
the geometry of the emitting region, the disk emissivity, and the disk's 
inclination angle with respect to the line of sight of the distant observer.
In the Kerr spacetime, the only relevant parameter of the background 
geometry is the spin parameter $a_*$, while $M$ sets the length of the 
system, without affecting the shape of the line. In those sources for which 
there is indication that the line is mainly emitted close to the compact 
object, the emission region may be thought to range from the radius of 
the ISCO, $r_{\rm in} = r_{\rm ISCO}$, to some outer radius $r_{\rm out}$. 
However, even more complicated geometries are possible. In principle, 
the disk emissivity could be theoretically calculated. In practice, that is 
not feasible at present. The simplest choice is an intensity profile 
$I_{\rm e} \propto r^{\alpha}$ with index $\alpha < 0$ to be determined 
during the fitting procedure. The fourth parameter is the inclination of the 
disk with respect to the line of sight of the distant observer, $i$. The 
dependence of the line profile on $a_*$, $i$, $\alpha$, and $r_{\rm out}$ 
in the Kerr background has been analyzed in detail by many authors, 
starting with Ref.~\cite{fab89}. The line profile in non-Kerr backgrounds is 
discussed in~\cite{ka-cb-a,ka-cb-b,ka-jp,ka-dan}.

Roughly speaking, the calculation of the profile of the K$\alpha$ iron
line goes as follows. We want to compute the photon flux number 
density measured by a distant observer, which is still given by
\be
N_{E_{\rm obs}} &=& \frac{1}{E_{\rm obs}} 
\int I_{\rm obs}(E_{\rm obs}) d \Omega_{\rm obs} =
\frac{1}{E_{\rm obs}} \int w^3 I_{\rm e}(E_{\rm e}) 
d \Omega_{\rm obs} \, .
\ee
As the K$\alpha$ iron line is intrinsically narrow in frequency, we 
can assume that the disk emission is monochromatic (the rest frame energy 
is $E_{\rm{K}\alpha} = 6.4$~keV) and isotropic with a power-law radial 
profile:
\be
I_{\rm e}(E_{\rm e}) \propto \delta (E_{\rm e} - E_{\rm{K}\alpha}) r^{\alpha} \, .
\ee
The calculation of $w$ is the same of the one for the disk's thermal spectrum,
as well as the calculation of the photon trajectories, from the point of emission 
in the disk to the image plane of the distant observer. More details can be found 
in Ref.~\cite{ka-cb-a}.

\section{Testing the spacetime geometry around black hole candidates \label{s-test}}

In both the continuum-fitting method and the analysis of the K$\alpha$ iron
line, the background geometry sets the inner edge of the disk, which is supposed
to be at the ISCO radius. That is a very important ingredient in the calculation
of the properties of the radiation. In the disk's spectrum, the inner edge of the 
disk determines the radiative efficiency of the accretion process, and thus
the high energy cut-off of the spectrum, which is the key-point in the measurement
of the spin. In the profile of the iron line, the inner edge of the disk determines
the low energy tail of the line. The geometry of the spacetime 
enters also the calculations of the redshift factor $w$, of the propagation of
the photons from the disk to the image plane of the observer, and of the 
time-averaged energy flux of the disk (for the thermal spectrum of the disk).
If we want to test the Kerr nature of an astrophysical BH candidate, it is
convenient to consider a general spacetime in which the central object is
described by a mass $M$, spin parameter $a_*$, and one (or more)
``deformation paramater(s)''. The latter measure possible deviations from the
Kerr solution, which must be recovered when all the deformation parameters
vanish. The strategy is thus to calculate the properties of the radiation emitted 
by the gas in the accretion disk in these more general backgrounds and then
fit the observational data of the source to find the allowed values of the
spin and of the deformation parameters. If the observations require vanishing
deformation parameters, the compact object is a Kerr BH. 
If they demand non-vanishing deformation parameters,
astrophysical BH candidates are not the Kerr BH of general relativity and new
physics is necessary. In general, however, the result is that observations
allows both the possibility of a Kerr BH with a certain spin parameter and
non-Kerr objects with different spin parameters.

\begin{figure}
\begin{center}
\hspace{-1.5cm}
\includegraphics[type=pdf,ext=.pdf,read=.pdf,width=10cm]{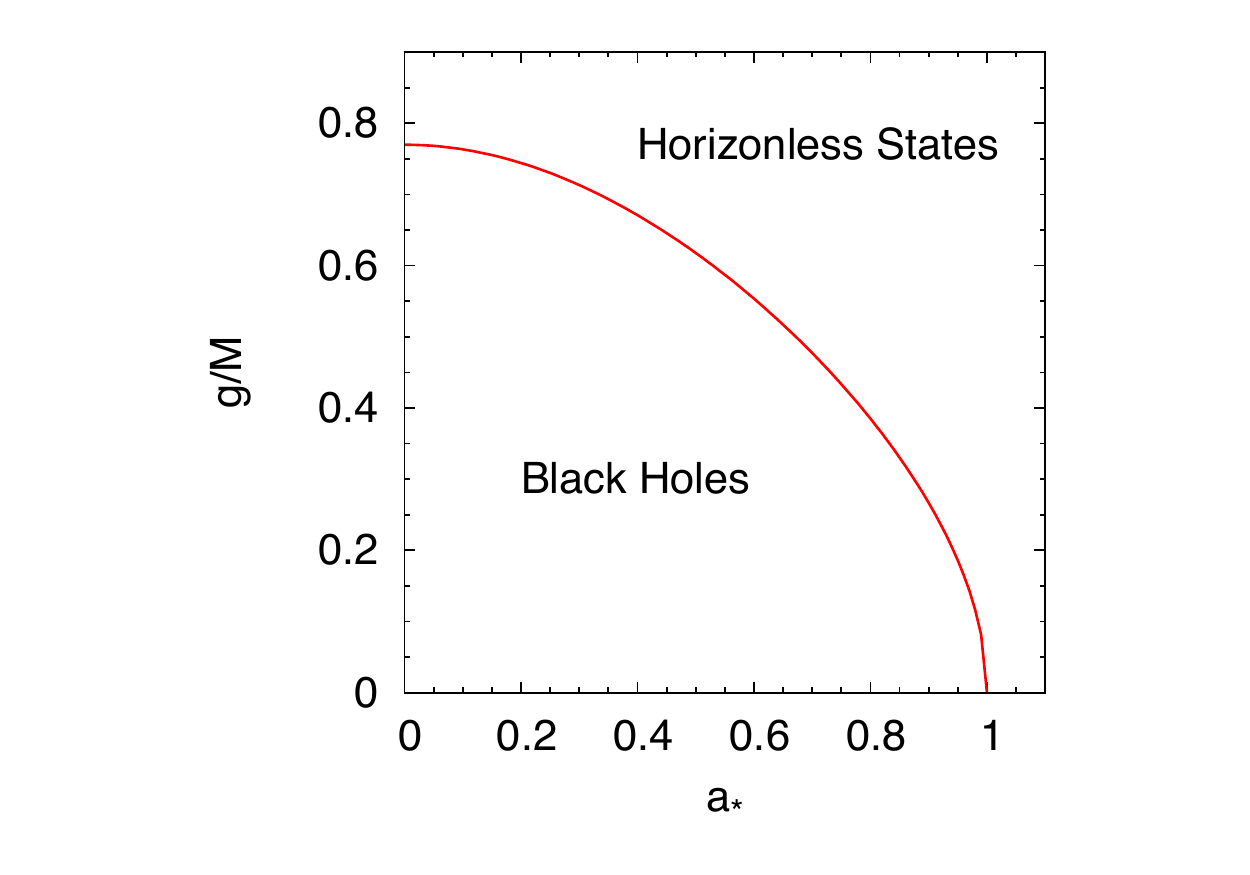}
\end{center}
\vspace{-0.5cm}
\caption{Rotating Bardeen metric. The red solid line separates the region
with BHs from the one with horizonless objects.}
\label{fig1}
\end{figure}

As test-metric, let us consider the rotating Bardeen black hole solution~\cite{regular}:  
\be\label{eq-bar}
ds^2 &=& - \left(1 - \frac{2 m r}{\Sigma}\right) dt^2 
- \frac{4 a m r \sin^2\theta}{\Sigma} dt d\phi
+ \frac{\Sigma}{\Delta} dr^2 + \Sigma d\theta^2 
+ \nonumber\\ &&
+ \sin^2 \theta \left(r^2 + a^2 
+ \frac{2 a^2 m r \sin^2\theta}{\Sigma} \right) d\phi^2 \, ,
\ee
where $a = J/M$, $\Sigma = r^2 + a^2 \cos^2\theta$, $\Delta = r^2 - 2 m r + a^2$, and
\be\label{eq-m}
m = \frac{M r^3}{\left(r^2 + g^2\right)^{3/2}} \, .
\ee
The deformation parameter is the charge $g$ and the Kerr metric is recovered
when $g = 0$. The radius of the event horizon is given by the larger root of $\Delta = 0$. 
If the equation $\Delta = 0$ has no solutions, the object is not a BH but a configuration
without horizon. The regions of BHs and horizonless states on the plane $(a_*, g/M)$
are shown in Fig.~\ref{fig1}. In what follows, I will restrict the attention to the BH
region: even if they can be created~\cite{z2}, the horizonless states are likely very 
unstable objects with a short lifetime due to the ergoregion instability.

\begin{figure}
\begin{center}
\hspace{-1cm}
\includegraphics[type=pdf,ext=.pdf,read=.pdf,width=9cm]{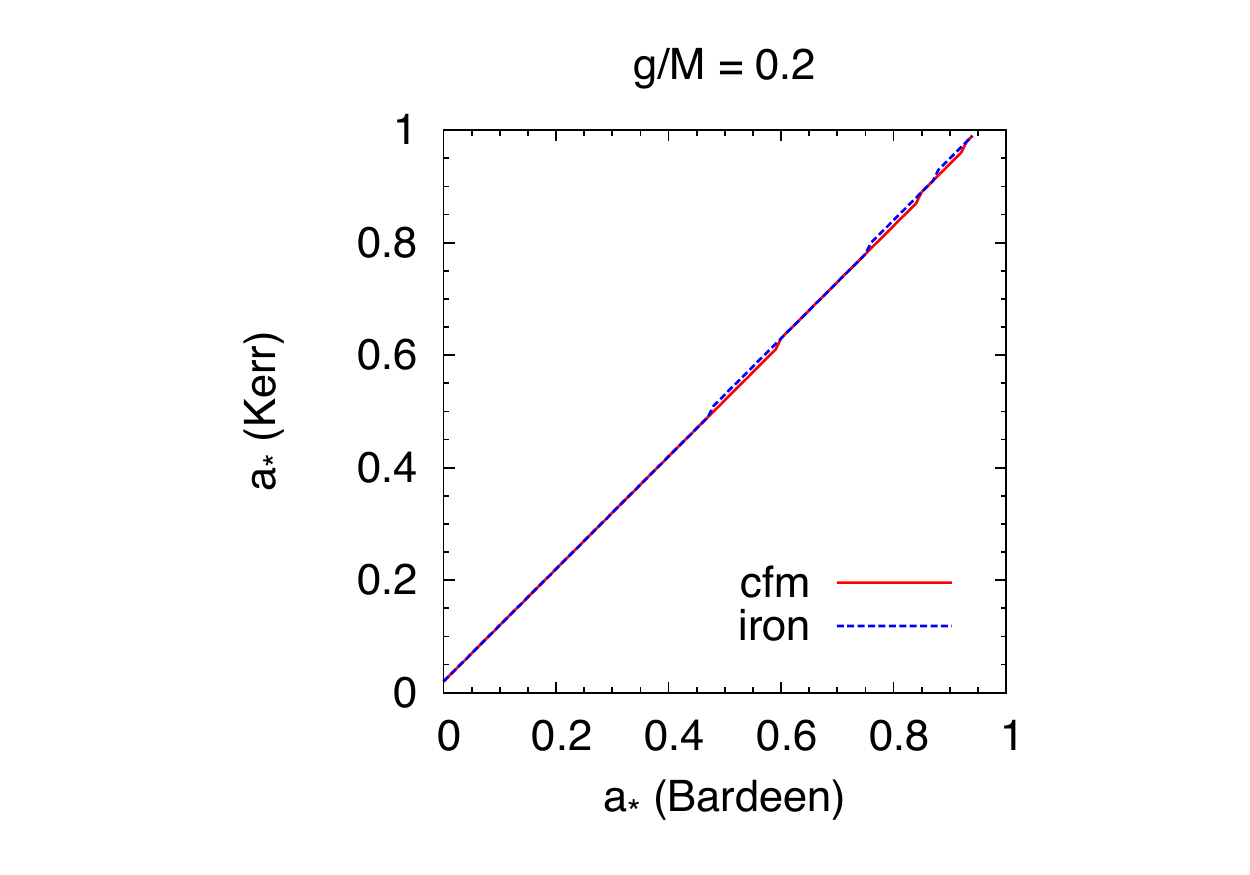} \hspace{-2cm}
\includegraphics[type=pdf,ext=.pdf,read=.pdf,width=9cm]{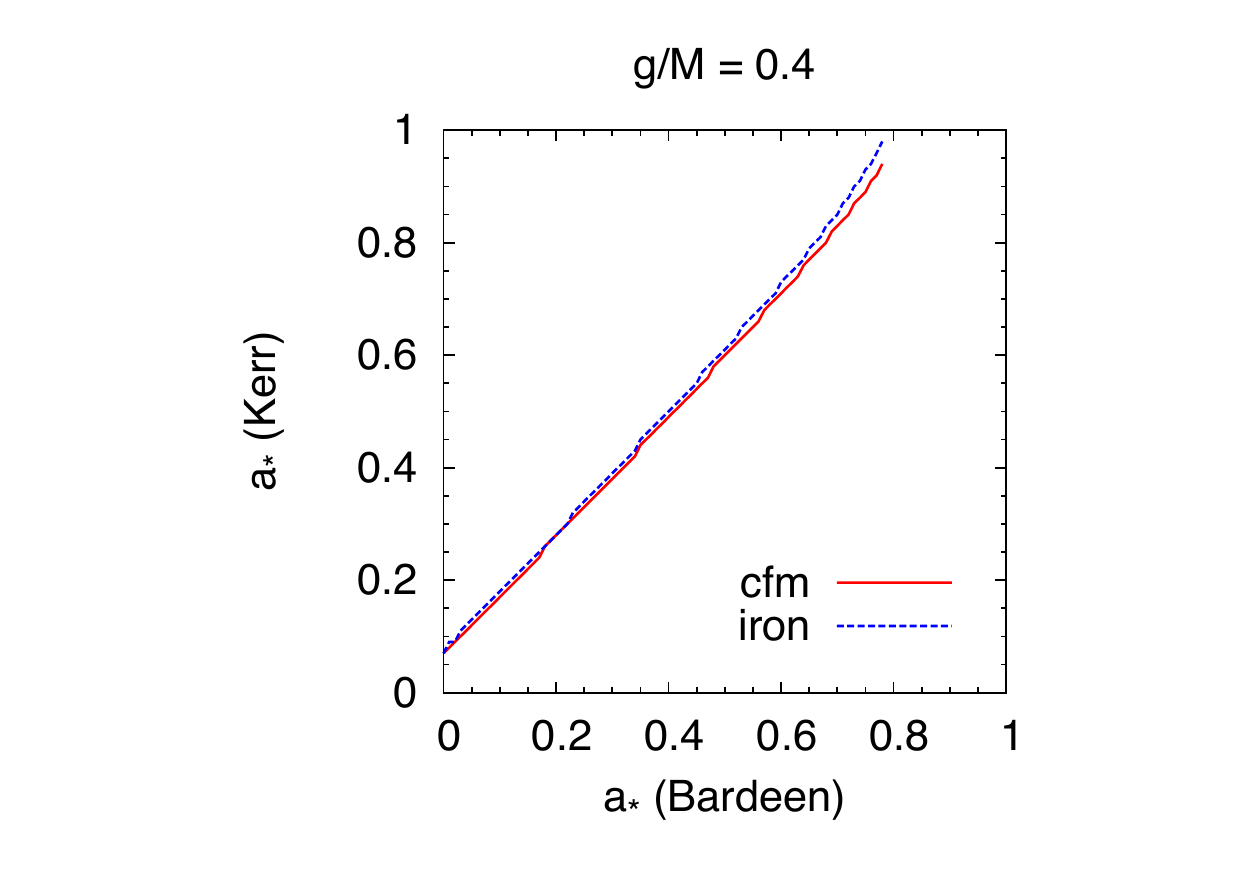} \\
\hspace{-1cm}
\includegraphics[type=pdf,ext=.pdf,read=.pdf,width=9cm]{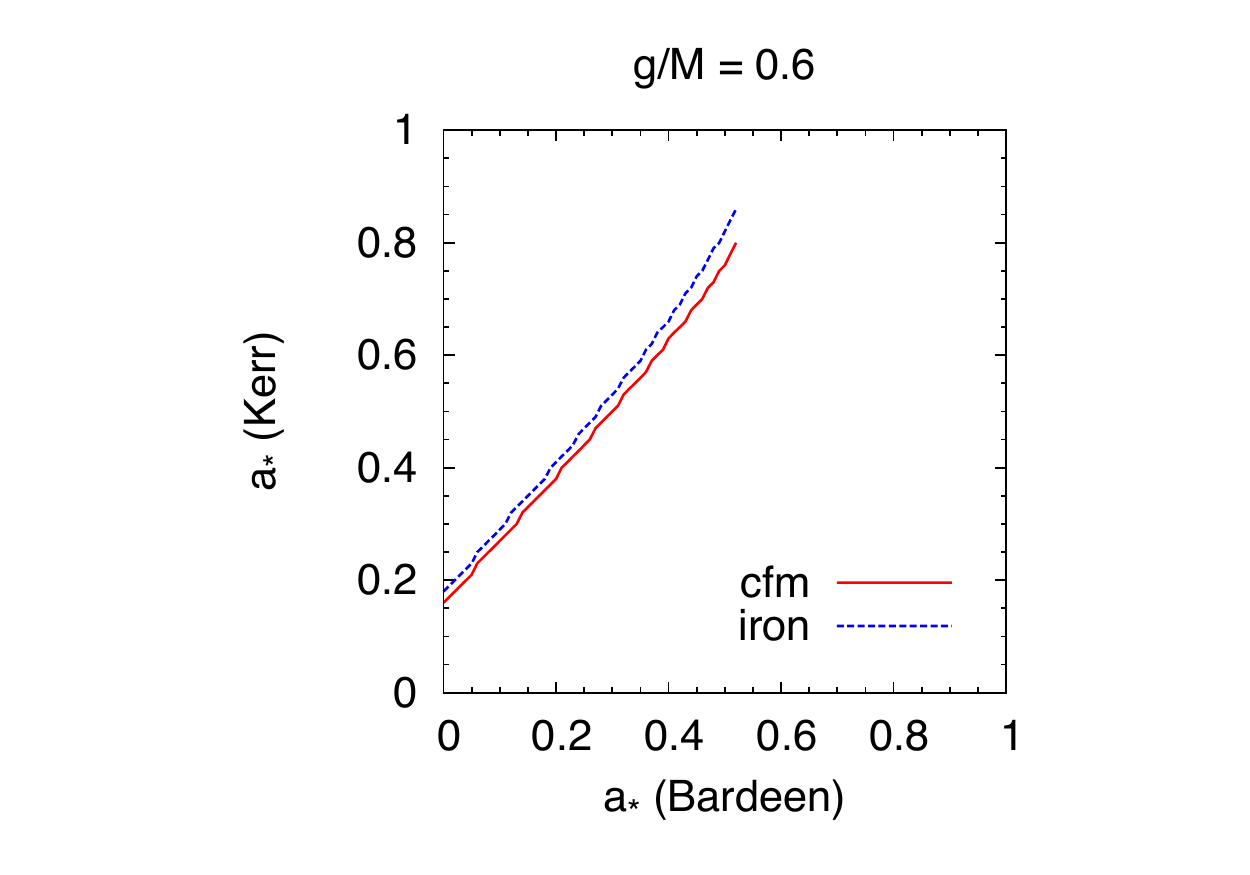} \hspace{-2cm}
\includegraphics[type=pdf,ext=.pdf,read=.pdf,width=9cm]{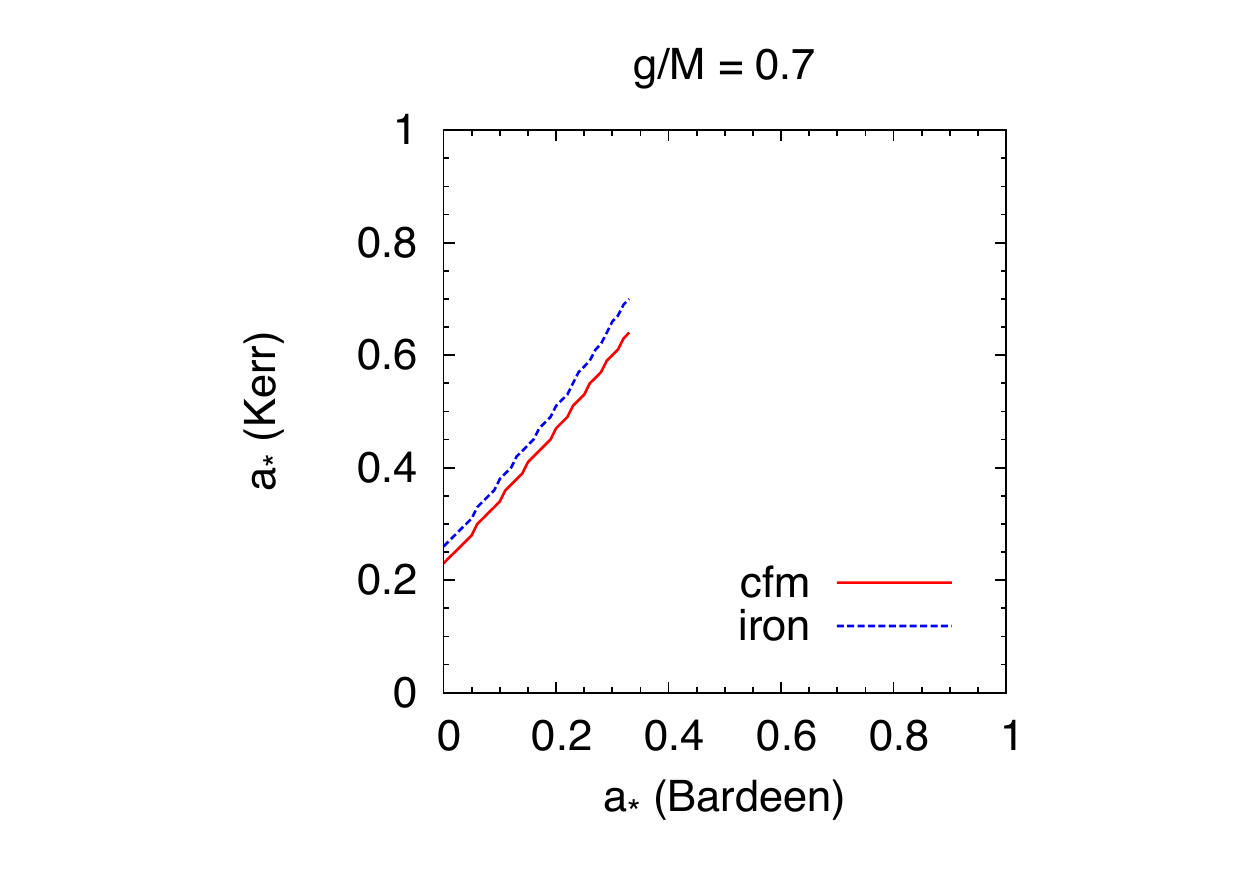}
\end{center}
\vspace{-0.5cm}
\caption{Kerr spin measurement via the continuum-fitting method (red solid lines) 
and the iron line analysis (blue dashed lines) in the case the astrophysical source
is a rotating Bardeen black hole with $g/M = 0.2$ (top left panel), 0.4 (top right panel), 
0.6 (bottom left panel), and 0.7 (bottom right panel). Even if for high values of $g/M$
the difference between the actual value of the spin parameter and the one inferred
assuming the Kerr background can be significant, the two techniques provide quite
similar estimates. See the text for details.}
\label{fig2}
\end{figure}

Let us now consider the possibility that an astrophysical BH candidate is a
Bardeen BH and that we want to measure the spin parameter of this object
with the continuum-fitting method and the K$\alpha$ iron line analysis,
assuming that the object is a Kerr BH. In this case, we can use an approach
similar to the one discussed in Ref.~\cite{ka-cb-a} and define the reduced 
$\chi^2$, respectively for the disk's thermal spectrum and the iron line profile:
\be
\label{eq-chi2-cfm}
\chi^2_{\rm disk, \; red} (a_*^{\rm Kerr}, i) &=&
\frac{1}{n} \sum_{i = 1}^{n} \frac{\left[N_{{\rm disk } \; i}^{\rm Kerr} 
(a_*^{\rm Kerr}, i) - N_{{\rm disk} \; i}^{\rm B} (a_*^{\rm B}, g/M, i^{\rm B})
\right]^2}{\sigma^2_i} \, , \\
\chi^2_{\rm iron, \: red} (a_*^{\rm Kerr}, i)
&=& \frac{1}{n} \sum_{i = 1}^{n} \frac{\left[N_{{\rm iron} \; i}^{\rm Kerr} 
(a_*^{\rm Kerr}, i) - N_{{\rm iron}\; i}^{\rm B}
(a_*^{\rm B}, g/M, i^{\rm B}) 
\right]^2}{\sigma^2_i} \, ,
\label{eq-chi2-ka}
\ee 
where the summation is performed over $n$ sampling energies $E_i$ and
$N_i^{\rm Kerr}$ and $N_i^{\rm B}$ are the photon fluxes (normalized photon 
fluxes in the case of the iron line profile) in the energy bin $[E_i,E_i+\Delta E]$,
respectively for the Kerr and the Bardeen metric. The error $\sigma_i$ is 
assumed to be 15\% the photon flux $N_i^{\rm B}$, which is roughly the 
accuracy of current observations in the best situations. For the analysis, I take 
$i^{\rm B} = 45^\circ$ and $43^\circ < i < 47^\circ$, assuming that the inclination
angle of the disk is known from independent observations with an accuracy
$\pm 2^\circ$. However, the final result is not sensitive to the exact choices of 
$i^{\rm B}$ and $\Delta i$. For the disk's spectrum, all the calculations are done with 
$M = 10$~$M_\odot$ and $\dot{M} = 2 \cdot 10^{18}$~g~s$^{-1}$. The mass
accretion rate should actually be inferred during the fitting procedure, but here
we can assume to be known as its determination is independent of the spin
measurement. In the case of the iron line, all the calculations are done
with an intensity profile index $\alpha = -3$ and an outer radius 
$r_{\rm out} = r_{\rm in} + 100 \, M$.

At this point, one can consider a specific value of $a_*^{\rm B}$ and $g/M$ 
and find the minimum of the reduced $\chi^2$, thus obtaining the measurement
of the Kerr spin parameter. The results are reported in Fig.~\ref{fig2} for
$g/M = 0.2$ (left top panel), 0.4 (right top panel), 0.6 (left bottom panel), and
0.7 (right bottom panel), respectively for the disk's thermal spectrum (red solid
lines) and the iron line (blue dashed lines). Let us note a few important things.
First, in all these cases the fit with the Kerr metric is good, in the sense that
the minimum of $\chi^2_{\rm red}$ is $\ll 1$, for both the continuum-fitting method
and the K$\alpha$ iron line analysis. So, the properties of these spectra
are really indistinguishable from the one of a disk around a Kerr BH. 
Second, the two techniques provide quite similar results. For $g/M = 0.2$,
the discrepancy is always $\Delta a^{\rm Kerr}_* \lesssim 0.01$. For $g/M = 0.7$,
the difference between the spin measurements of the two approaches ranges
from $\Delta a_*^{\rm Kerr} \approx 0.03$, for non-rotating and slow-rotating 
Bardeen BHs, to $\Delta a^{\rm Kerr}_* \approx 0.06$, for near extremal objects.
The  spin range is $a_*^{\rm B} < 0.95$ for $g/M = 0.2$, $a_*^{\rm B} < 0.79$ 
for $g/M = 0.4$, $a_*^{\rm B} < 0.53$ for $g/M = 0.6$, and $a_*^{\rm B} < 0.34$ 
for $g/M = 0.7$ because for higher spins there are no BHs, but some kind
of horizonless states (like the Kerr metric with $a^{\rm Kerr}_* > 1$, but without 
central singularity). Because of the ergoregion instability, fast-rotating compact
objects without an event horizon are expected to be very unstable, and so not
good BH candidates. However, even if they were stable, their features would be
very different and thus these objects would be easy to distinguish from Kerr BHs.

\begin{figure}
\begin{center}
\hspace{-1cm}
\includegraphics[type=pdf,ext=.pdf,read=.pdf,width=9cm]{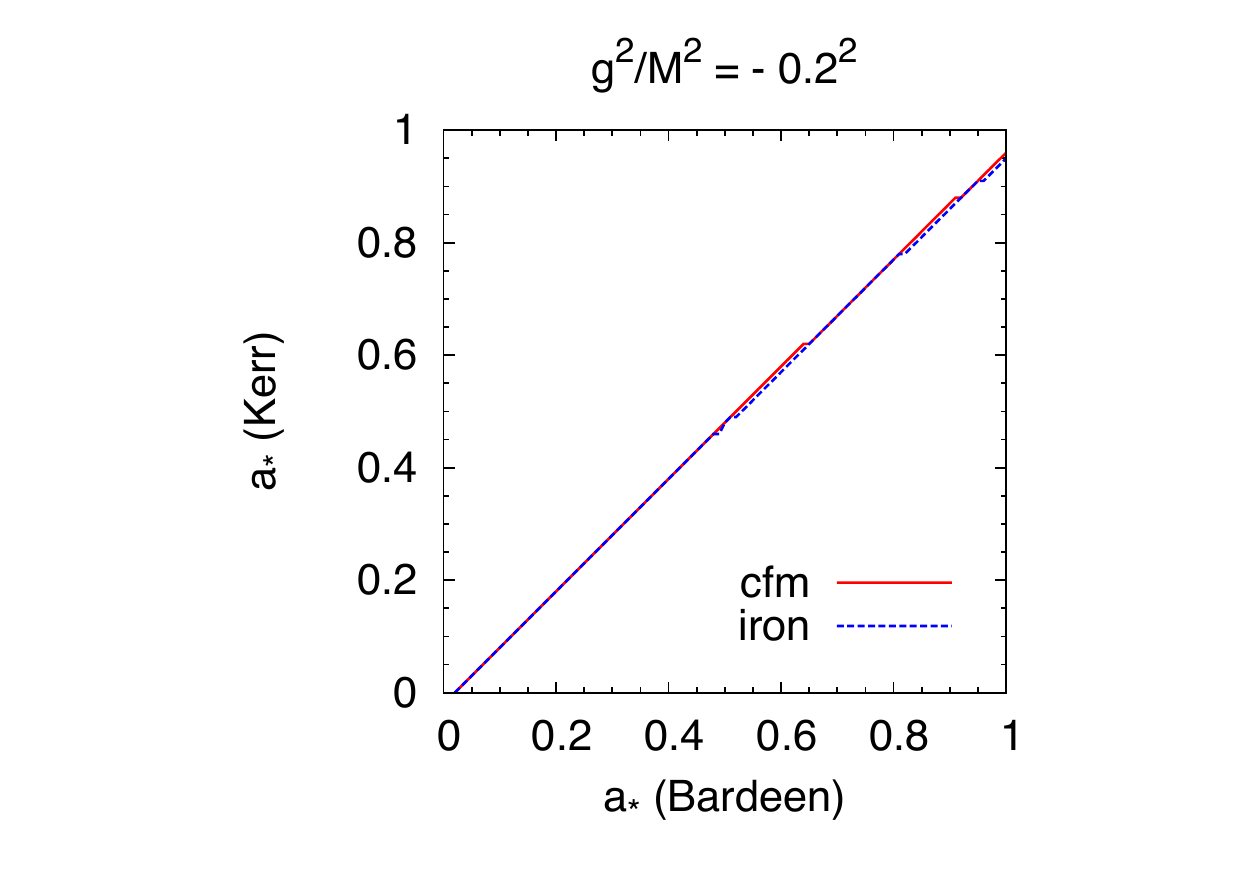} \hspace{-2cm}
\includegraphics[type=pdf,ext=.pdf,read=.pdf,width=9cm]{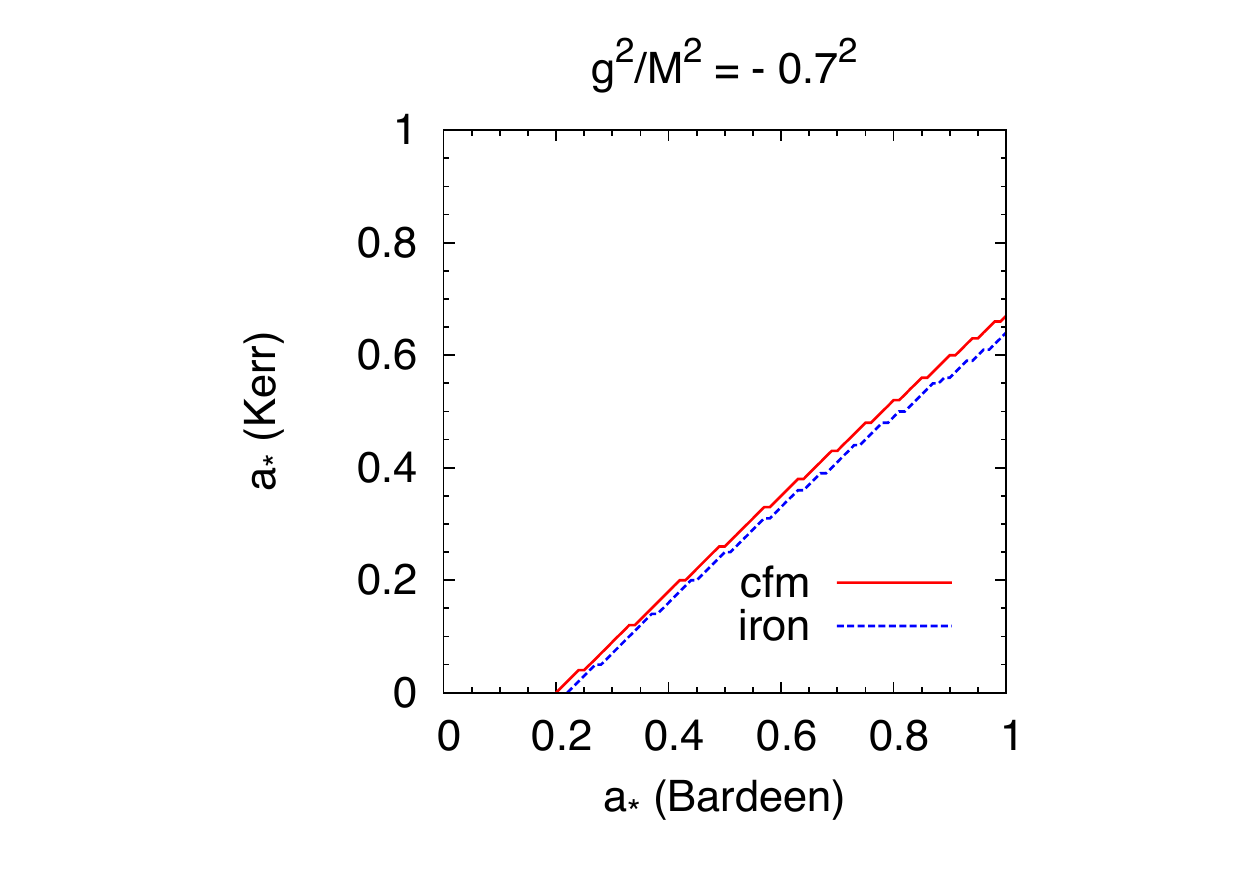}
\end{center}
\vspace{-0.5cm}
\caption{As in Fig.~\ref{fig2}, for $g^2/M^2 = -0.2^2$ (left panel) and $-0.7^2$ 
(right panel). See the text for details.}
\label{fig3}
\end{figure}

Fig.~\ref{fig2} shows that the inferred Kerr spin parameter is always higher than
its actual value. This is only the consequence of the specific form of the Bardeen
metric, in which the gravitational force is weaker than the one around a Kerr
BH with the same spin. To have a stronger gravitational field, we just need that
$g^2$ in Eq.~\eqref{eq-m} is negative. While such a possibility is against the
original motivation of the Bardeen metric, i.e. to have a singularity free BH solution,
we can surely consider this case in our approach. When $g^2 < 0$, the measurement
of the Kerr spin provides a value lower than the actual spin parameter of the object,
as shown in Fig.~\ref{fig3}. However, the difference between the results obtained
via the continuum-fitting and the iron line methods are still quite similar.

\section{Discussion \label{s-d}}

The thermal spectrum of a geometrically thin disk is determined by the mass
of the compact object, its distance from us, the inclination angle of the disk with
respect to the observer's line of sight, the mass accretion rate, and the spin (assuming
the Kerr background; otherwise the spin and the deformation parameter(s) if
we want to test the nature of the BH candidate). The first three parameters
(mass, distance, viewing angle) must be known from independent measurements.
One can then fit the soft X-ray component of the spectrum  of the source and 
obtain the mass accretion rate and the background parameters. It is clear
that the source must be in the high-soft state, when the component from the
disk is dominant. In the case of the iron line analysis, the mass and the distance
of the compact object do not enter the calculations of the line profile, while the
viewing angle can be inferred during the fitting procedure and it is thus not an
input parameter. Good data of the iron line usually require that the source is in 
the low-hard state. From this considerations, it follows that it is not obvious to
be able to measure the spin parameter with both the continuum-fitting method
and the K$\alpha$ iron analysis of the same source, as we need different
conditions and observations.

In Ref.~\cite{xte}, the authors consider the BH candidate in XTE~J1550-564 
and they measure the Kerr spin parameter of this object with the analysis of
the disk's thermal spectrum and of the iron line profile. They find
\be
a_*^{\rm Kerr} &=& 0.34^{+ 0.23}_{- 0.28} \;\; (68\% \; {\rm C.L.}) \, , \;\;
0.34^{+ 0.37}_{- 0.45} \;\; (90\% \; {\rm C.L.}) \quad ({\rm disk's \; spectrum}) \, , \\
a_*^{\rm Kerr} &=& 0.55^{+ 0.05}_{- 0.11} \;\; (68\% \; {\rm C.L.}) \, , \;\;
0.55^{+ 0.15}_{- 0.22} \;\; (90\% \; {\rm C.L.}) \quad ({\rm iron \; line}) \, .
\ee
The two measurement are consistent, but the uncertainty is quite large.
From the results of the previous section, it is clear that to distinguish a Kerr
BH from a Bardeen one with the measurement of the Kerr spin we need,
in the most optimistic situation of a fast-rotating object with large $g/M$, 
measurements with an accuracy at the level of $\Delta a_*^{\rm Kerr} \approx 0.02$, 
much better than what we can do now and presumably out of reach even 
in the future.

Are we really so far from being able to test the Kerr nature of BH candidates? 
Actually, if we consider a different test-metric, our chances may not be so bad.
To be more specific, let us replace the rotating Bardeen solution with the 
Johannsen-Psaltis (JP) metric, proposed in Ref.~\cite{jp-m} explicitly to be
used as a metric to test the Kerr BH hypothesis and potentially capable of
describing non-Kerr BHs in a putative alternative theory of gravity. In its simplest 
version, the line element in Boyer-Lindquist coordinates reads
\be\label{eq-jp}
ds^2 &=& - \left(1 - \frac{2 M r}{\Sigma}\right) \left(1 + h\right) dt^2
- \frac{4 a M r \sin^2\theta}{\Sigma} \left(1 + h\right) dtd\phi
+ \frac{\Sigma \left(1 + h\right)}{\Delta + a^2 h \sin^2 \theta} dr^2
+ \nonumber\\ &&
+ \Sigma d\theta^2 + \left[ \left(r^2 + a^2 +
\frac{2 a^2 M r \sin^2\theta}{\Sigma}\right) \sin^2\theta +
\frac{a^2 (\Sigma + 2 M r) \sin^4\theta}{\Sigma} h \right] d\phi^2 \, ,
\ee
where now $\Delta = r^2 - 2 M r + a^2$ and 
\be
h = \frac{\epsilon_3 M^3 r}{\Sigma^2} \, .
\ee
Here $\epsilon_3$ is the deformation parameter. The compact object is more 
prolate (oblate) than a Kerr BH for $\epsilon_3 > 0$ ($\epsilon_3 < 0$); when 
$\epsilon_3 = 0$, we recover the Kerr solution.

\begin{figure}
\begin{center}
\hspace{-1cm}
\includegraphics[type=pdf,ext=.pdf,read=.pdf,width=9cm]{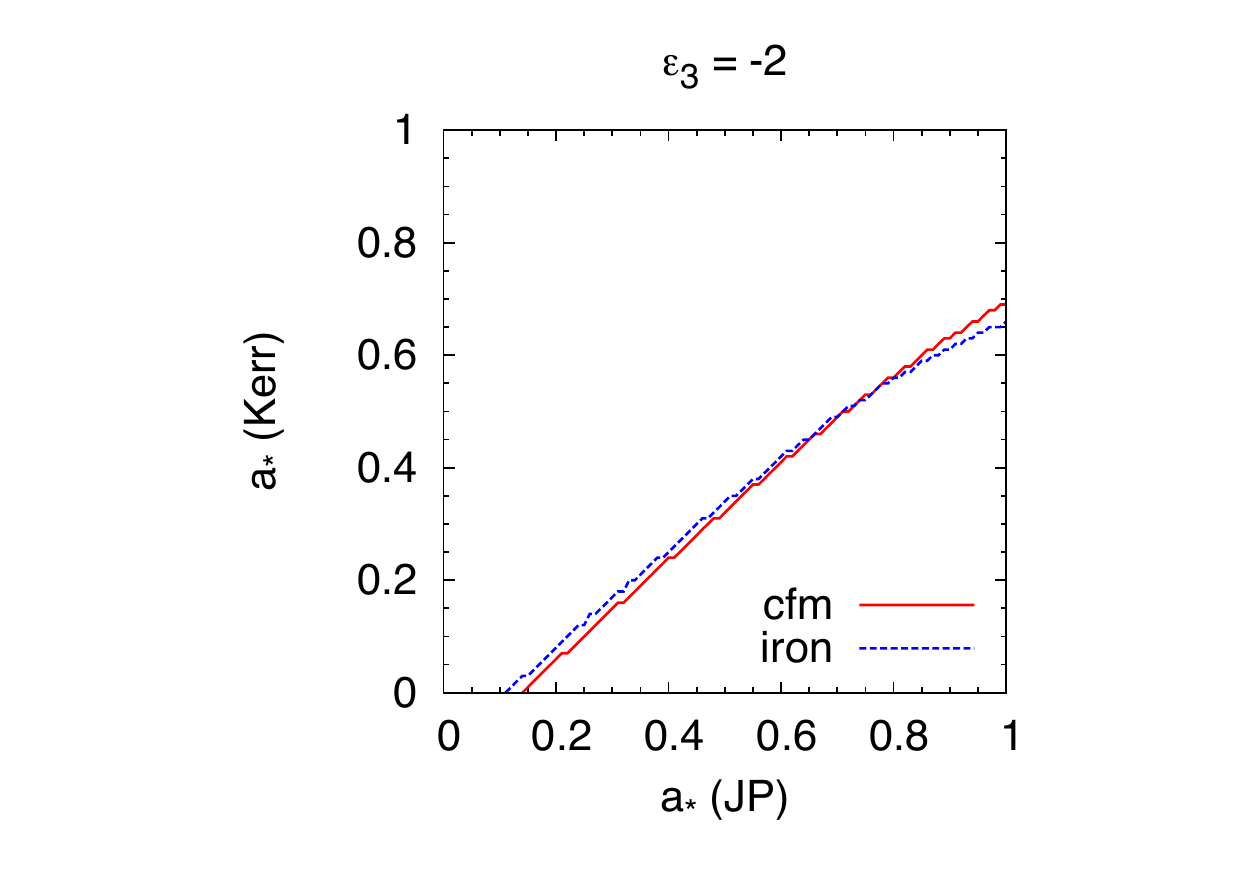} \hspace{-2cm}
\includegraphics[type=pdf,ext=.pdf,read=.pdf,width=9cm]{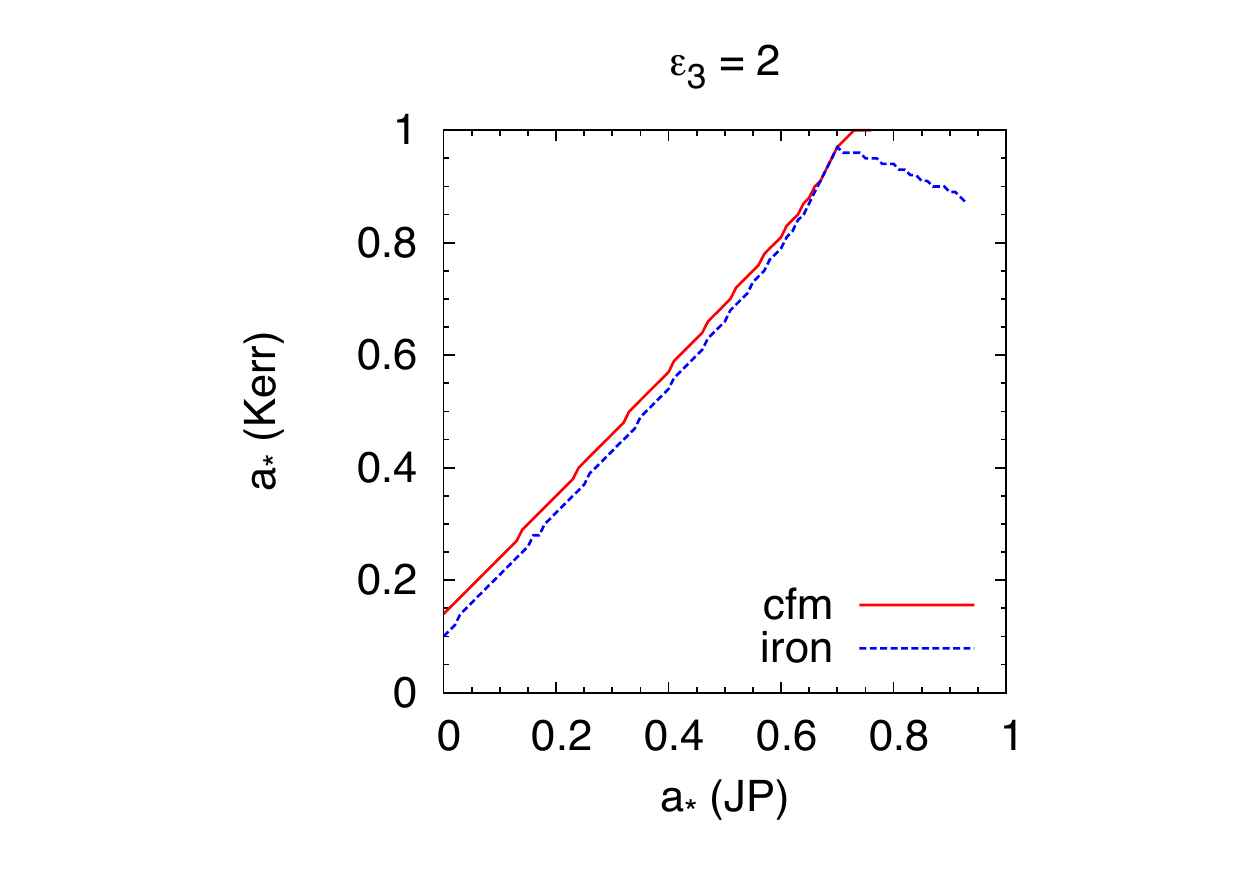} \\
\hspace{-1cm}
\includegraphics[type=pdf,ext=.pdf,read=.pdf,width=9cm]{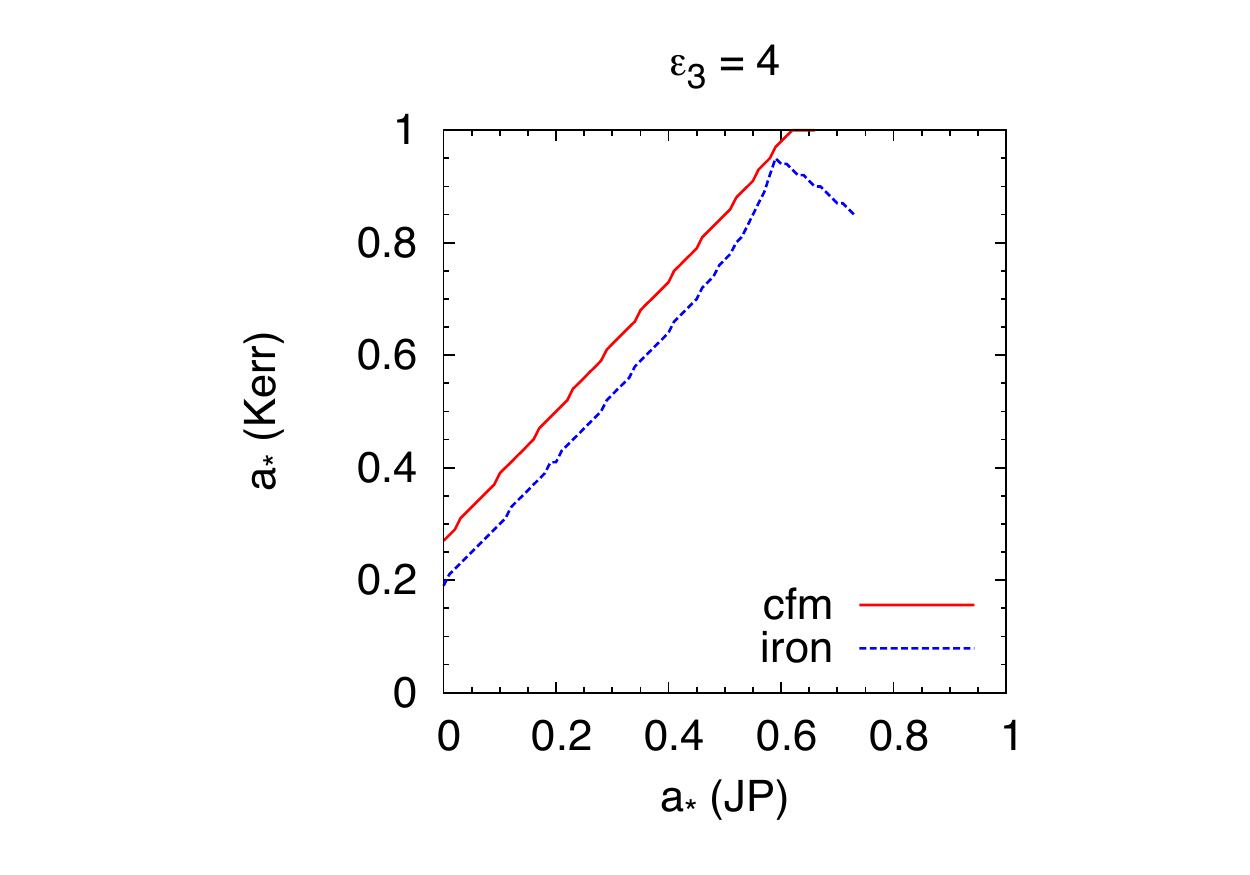} \hspace{-2cm}
\includegraphics[type=pdf,ext=.pdf,read=.pdf,width=9cm]{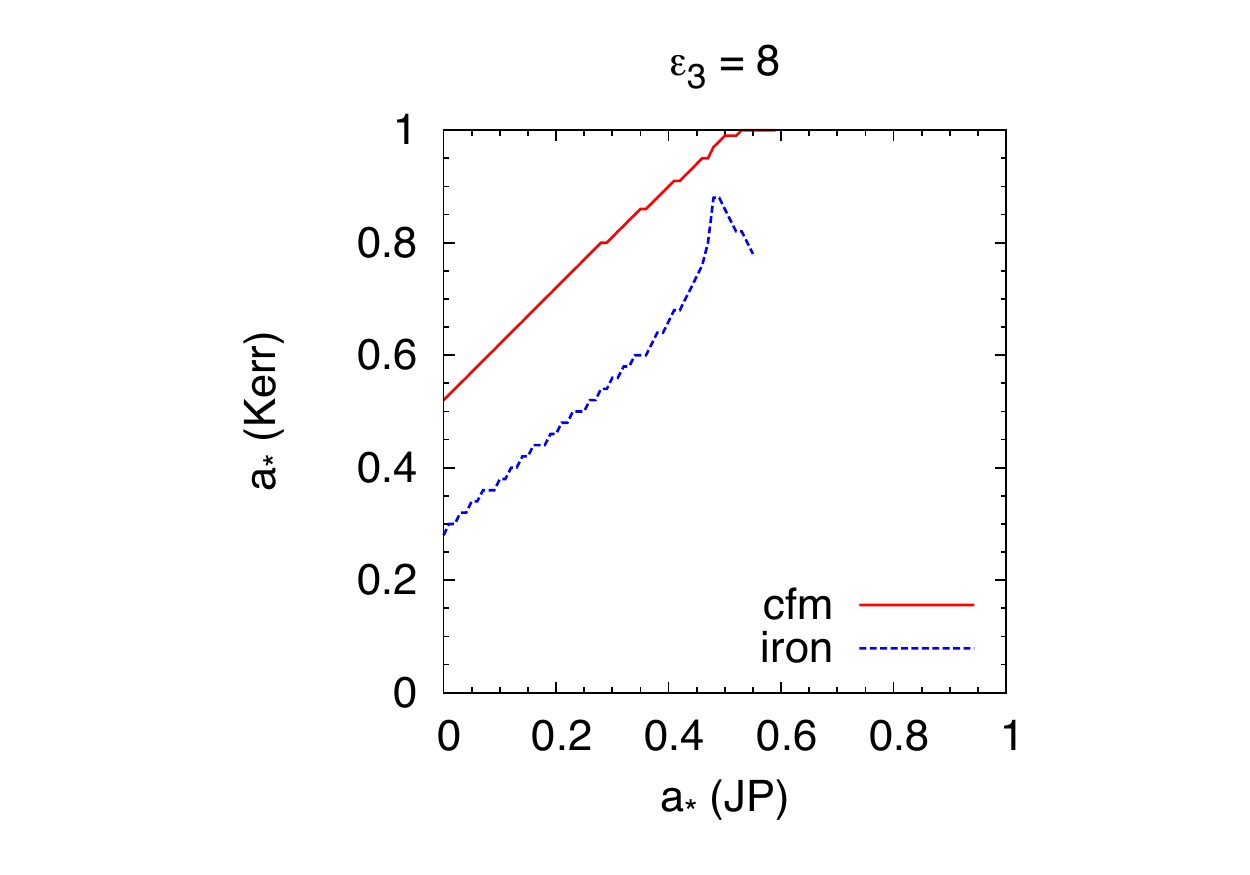}
\end{center}
\vspace{-0.5cm}
\caption{As in Fig.~\ref{fig2}, for the JP metric with $\epsilon_3 = -2$ (top left 
panel), 2 (top right panel), 4 (bottom left panel), and 8 (bottom right panel). 
See the text for details.}
\label{fig4}
\end{figure}

If we proceed as in the previous section and we compare the Kerr spin 
measurements via the continuum-fitting and the iron line methods, we find
that the discrepancy between the two techniques may be larger. In Fig.~\ref{fig4},
I show the cases $\epsilon_3 = -2$, 2, 4, and 8. Here it is not really clear the
maximum value of the spin parameter that makes sense to consider. In 
Fig.~\ref{fig4}, I considered spins from 0 to 1. The end of the lines
in the left panel of Fig.~\ref{fig4} is simply due that for higher values of the
spin parameter the fits become bad and the minimum of $\chi^2_{\rm red}$
exceeds 1. So, at this point the two metrics (JP and Kerr) become very
different and even a single measurement (continuum-fitting or iron line)
can check if the BH candidate is or is not a Kerr BH.
However, these cases with a bad fit are usually configurations
with naked singularities, whose physical meaning would be anyway 
questionable. In the case $\epsilon_3 = 8$, the discrepancy between the 
Kerr spin measurements of the two techniques is $\Delta a_*^{\rm Kerr} 
\approx 0.25$, quite independently of the exact value of the spin for 
slow-rotating and mid-rotating objects. While these results may suggest that
future X-ray data could really test the nature of a BH candidate, it is important
to stress a fundamental difference between the metric in Eq.~\eqref{eq-bar} 
and the one in~\eqref{eq-jp}. The Bardeen metric is an acceptable BH
solution in presence of exotic matter and presumably even in the case of
extensions of general relativity. The JP metric is obtained with an arbitrary
procedure and it is definitively not clear if such kinds of deformations from
the Kerr geometry can be obtained from a consistent theory. It is found by
starting from a deformed Schwarzschild solution and it is transformed into a 
rotating solution with the Newman-Janis algorithm. However, the initial 
deformed non-rotating metric has $g_{tt} \neq -1/g_{rr}$, which is not the
case one would expect for a BH. More importantly, the transformation to
get the metric in Boyer-Lindquist coordinates is not a valid coordinate
transformation~\cite{regular} (see also the discussion in Ref.~\cite{maa}).
The result is that the metric has some quite pathological features~\cite{pat}.
So, the interpretation of these results remain ambiguous, but at least they
show that a discrepancy between the continuum-fitting method and the
iron line analysis is not necessarily tiny and it depends on the specific 
background metric.

\begin{figure*}
\begin{center}
\hspace{-1.5cm}
\includegraphics[type=pdf,ext=.pdf,read=.pdf,width=9cm]{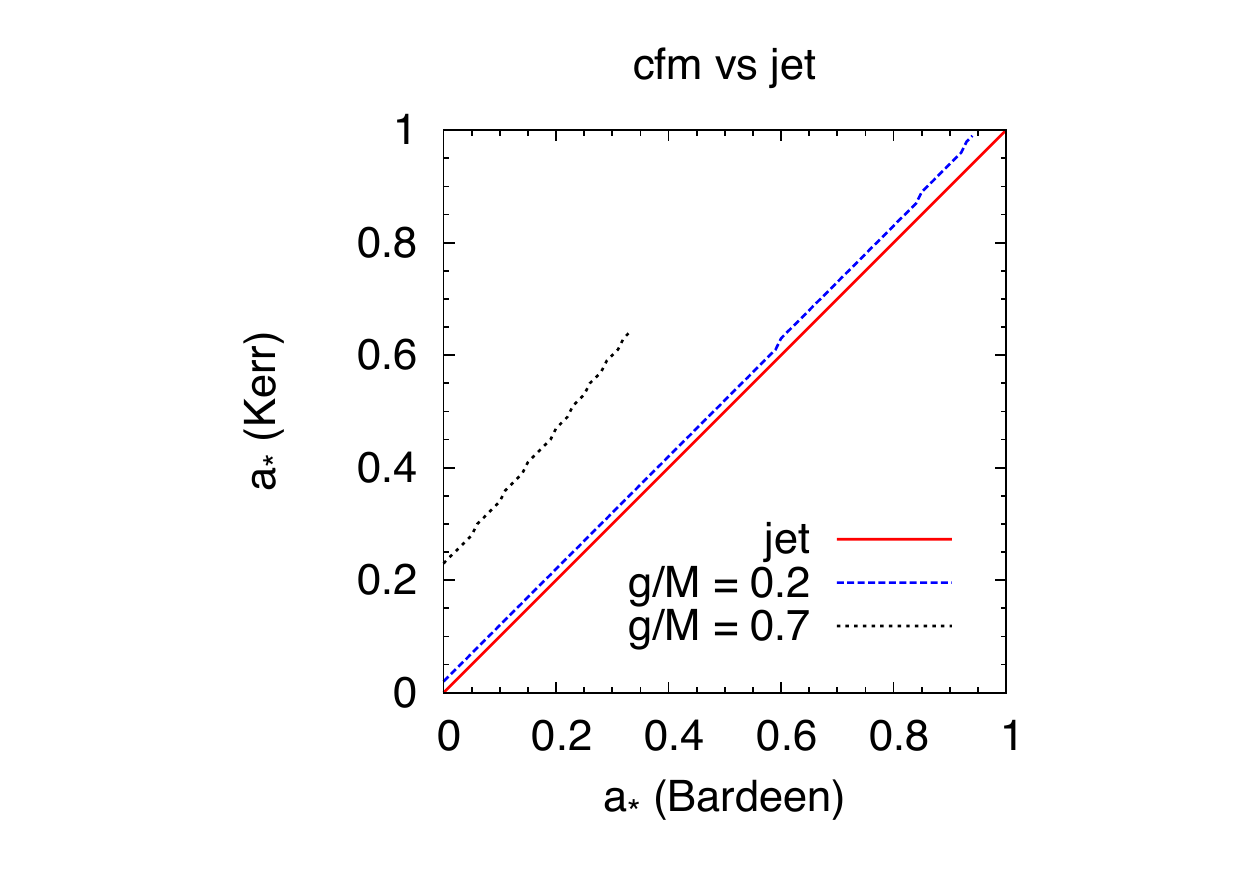}
\end{center}
\vspace{-0.3cm}
\caption{Comparison of the Kerr spin measurements via the jet power
and the continuum-fitting method in the case the astrophysical source is 
a rotating Bardeen BH. The jet power approach may really measure
the spin parameter of the object, independently of its actual nature.
See the text for details.}
\label{fig5}
\end{figure*}

Finally, one can also consider the possibility of using other approaches
to combine with the measurements of the disk's spectrum or of the iron
line profile. Unfortunately, however, other methods are not yet mature
to probe the spacetime geometry around BH candidates~\cite{bh1}.
The estimate of the radiative efficiency has been applied to non-Kerr
backgrounds in Refs.~\cite{radeff-a,radeff-b,radeff-c}, but it can only provide
a quite rough estimate of possible deviations from the Kerr solution.
Quasi-periodic oscillations seen in the X-ray spectrum of stellar-mass BH 
candidates may potentially be a quite powerful tool, but at present the exact
mechanism responsible for these features is not known and different
models provide different results~\cite{qpo}. An accurate observation of
the boundary of the ``shadow'' of BH candidates is also a very interesting
possibility, but it requires the use of sub-mm wavelength and
the necessary precision to test the Kerr metric may be out of reach 
for the near future~\cite{shadow-a,shadow-b,shadow-c,shadow-d,shadow-e}. 
Recently, there has been some interest in the possibility of measuring the
spin parameter (or both the spin and the deformation parameters) from
the estimate of the jet power~\cite{jet-si-a,jet-si-b,jet-cb-a,jet-cb-b}. The basic 
idea is that, if some kind of jets are powered by the spin of the compact 
object, the power of the jet should be correlated to the value of the spin 
parameter. Such an approach to measure the spin of the compact object
should be really independent of the background geometry and therefore
its combination with either the continuum-fitting measurement or the iron line
one would have the capability of revealing a possible non-Kerr nature of the
BH candidate, at least for large deformations, as shown in Fig.~\ref{fig5} for 
the Bardeen metric with $g/M = 0.2$ and 0.7. The issue is however quite
controversial~\cite{jet-no-a,jet-no-b}, and, depending if one assumes that 
steady or transient jets are powered by the BH spin, current data would
suggest different conclusions~\cite{jet-cb-a,jet-cb-b}.

\section{Summary and conclusions \label{s-c}}

Astrophysical BH candidates are stellar-mass compact objects in X-ray binary
systems and super-massive dark bodies at the center of every normal galaxy.
They are thought to be the Kerr BH of general relativity, but the actual nature
of these objects has still to be verified. To test the Kerr BH hypothesis, we have
to probe the spacetime geometry around these compact objects and check if 
there are deviations from the predictions of general relativity. This can be
achieved by studying the properties of the radiation emitted by the gas in the
accretion disk. Today we have only two relatively robust techniques to probe
the spacetime geometry around astrophysical BH candidates: the continuum-fitting
method and the analysis of the relativistic iron line profile. If we assume that
these objects are Kerr BHs, both the techniques can provide an estimate of the 
spin parameter. If we relax the Kerr assumption and we try to test the 
nature of the compact object, it turns out that we can only constrain a certain
combination of the spin and of possible deviations from the Kerr solution. In 
other words, both the disk's thermal spectrum and the iron line profile of a Kerr
BH with spin parameter $a_*$ is very similar, and practically indistinguishable, 
from the ones of non-Kerr objects with different spin parameters. In particular,
observations allow larger and larger deviations from the Kerr metric for 
more and more different spins with respect to the one expected from a Kerr
BH.

In this paper, I have discussed the quite natural question whether the use of 
the two techniques for a specific object can break the strong correlation between 
the spin and the deformations. Astronomers may indeed be particularly interested to 
know if consistent measurements of the Kerr spin with the continuum-fitting 
method and the iron line analysis can be used to claim the Kerr nature of the 
BH candidate. The result of the present work is that the answer is not necessary 
positive: at least in some non-Kerr metrics, the continuum-fitting method and the
analysis of the iron line profile may provide very similar results and it is
eventually impossible to distinguish a Kerr BH from another object. So, we
should be careful about a possible degeneracy between the spin and
deviations from the Kerr background. The reason of this degeneracy is only 
partially due to the common assumption of the two techniques that the
inner edge of the disk is at the ISCO radius. Indeed, the combination of the
two methods may fix this problem in other backgrounds (the JP metric is at 
least an example, even if its physical meaning is less clear). The main reason
seems instead to come from the form of the Bardeen metric, which, however,
is quite natural from a theoretical point of view~\cite{regular}: such a metric
can be written in Boyer-Lindquist coordinates as the Kerr one with the mass
$M$ replaced by a mass function $m (r)$, going to $M$ at large radii.
Other approaches, like the estimate of the jet power, are interesting possibilities 
to be combine with the continuum-fitting method and the K$\alpha$ iron line 
analysis to test even this kind of metrics.


\begin{acknowledgments}
This work was supported by the Thousand Young Talents Program 
and Fudan University.
\end{acknowledgments}


\end{document}